# Limits to Poisson's ratio in isotropic materials

P.H. Mott and C.M. Roland

Chemistry Division, Code 6120, Naval Research Laboratory, Washington DC 20375-5342

*(September 8, 2009)*

A long-standing question is why Poisson's ratio $\nu$ nearly always exceeds 0.2 for isotropic materials, whereas classical elasticity predicts $\nu$ to be between $-1$ to $\frac{1}{2}$. We show that the roots of quadratic relations from classical elasticity divide $\nu$ into three possible ranges: $-1 < \nu \leq 0$, $0 \leq \nu \leq \frac{1}{5}$, and $\frac{1}{5} \leq \nu < \frac{1}{2}$. Since elastic properties are unique there can be only one valid set of roots, which must be $\frac{1}{5} \leq \nu < \frac{1}{2}$ for consistency with the behavior of real materials. Materials with Poisson's ratio outside of this range are rare, and tend to be either very hard (e.g., diamond, beryllium) or porous (e.g., auxetic foams); such substances have more complex behavior than can be described by classical elasticity. Thus, classical elasticity is inapplicable whenever $\nu < \frac{1}{5}$, and the use of the equations from classical elasticity for such materials is inappropriate.

## I. INTRODUCTION

Classical elasticity continues to serve, without revision, as the basis for stress and strain analysis in science, engineering, and technology. The theory describes the reversible, linear mechanical response of a continuum, which for isotropic materials reduces to two governing constants. It provides expressions between all elastic constants and predicts that Poisson's ratio, a material constant defined as

$$\nu = -\frac{\varepsilon_{22}}{\varepsilon_{11}}, \quad (1)$$

where $\varepsilon_{22}$ and $\varepsilon_{11}$ are the lateral and axial strain for an axially loaded specimen, is limited to the range $-1$ to $\frac{1}{2}$.[1] These bounds are cited often[2,3,4]; however, in practice isotropic materials almost never have $\nu$ lower than 0.2, a discrepancy that remains unexplained since development of the theory in the 19th century.

The isotropic, two-parameter theory was first verified by measurement of Poisson's ratio in steel and brass beams in bending,[5] and the early work was carried out on similar substances. Unfortunately, confirmation in ordinary materials has lead to its uncritical application in extraordinary materials. For example, in studies of fused quartz[6], diamond[7], and beryllium[8], expressions from classical elasticity were used to find the elastic constants from wave speed measurements. On the other hand, it has been shown that porous auxetic materials do not obey classical elasticity.[9] As far as we know, confirmation of classical elasticity in materials for which $\nu < 0.2$ is nonexistent.

In this work we show that the origin of this long-standing issue can be resolved by using the roots of quadratic formulae from the classical theory to divide Poisson's ratio into three possible ranges. It is emphasized that since $\nu$ is unique, only a single set of roots can be valid.

## II. BACKGROUND AND THEORY

Classical elasticity posits a quadratic strain energy function, derived from the first law of thermodynamics, to govern the elastic response. For an isotropic body this function is[1]

$$2W = (\lambda + 2\mu)V^2 + \mu\gamma, \qquad (2)$$

for an infinitesimal strain tensor $\varepsilon_{ij}$. Here $\lambda$ and $\mu$ are the Lamé constants, $V\,(= \varepsilon_{11} + \varepsilon_{22} + \varepsilon_{33})$ is the volume change, and $\gamma\,(= (\varepsilon_{23})^2 + (\varepsilon_{31})^2 + (\varepsilon_{12})^2 - 4\varepsilon_{22}\varepsilon_{33} - 4\varepsilon_{33}\varepsilon_{11} - 4\varepsilon_{11}\varepsilon_{22})$ is the shear distortion. Differentiation of $2W$ with respect to $\varepsilon_{ij}$ defines the stress tensors $\sigma_{ij}$ to give the constitutive stress-strain relations; i.e., Hooke's law. Material elastic properties are measured in terms of the shear modulus $G\,(=\mu)$, Young's modulus $E$, and bulk modulus $B$, which are found from $\lambda$ and $\mu$ by applying the respective geometry to eq. 2. The well-known relations between any three elastic constants are derived from eq. 2 and are listed in standard texts.[10]

Thermodynamic stability requires that $G$, $E$, and $B$ are positive, finite, and non-zero; thus from [1,10]

$$B = \frac{2(1+\nu)}{3(1-2\nu)}G, \qquad (3)$$

it follows that $-1 < \nu < \tfrac{1}{2}$, which are the classical bounds to Poisson's ratio. Further limits to $\nu$ are obtained as follows.

In sound propagation the longitudinal modulus governing the compression wave speed is

$$M = \frac{\sigma_{11}}{\varepsilon_{11}}, \qquad (4)$$

where all other strains ($\varepsilon_{12}$, $\varepsilon_{33}$ etc.) equal zero. The longitudinal modulus is related to the bulk and shear moduli by $M = B + \tfrac{4}{3}G$; since $B$ and $G$ must be positive, $M$ must also be positive. Young's modulus as a function of $\nu$ and $M$ is

$$E = \frac{(1-2\nu)(1+\nu)}{1-\nu}M, \qquad (5)$$

which may be solved by the quadratic formula as

$$v = \frac{1}{4}\left[\frac{E}{M} - 1 \pm \left(\frac{E^2}{M^2} - 10\frac{E}{M} + 9\right)^{1/2}\right]. \tag{6}$$

The expression inside the square root can be factored into $(E/M - 9)(E/M - 1)$, so the square root is real only when $E/M \leq 1$ or $E/M \geq 9$. The stability requirements $E, M > 0$ and $v < \frac{1}{2}$ further restrict the range to $0 < E/M \leq 1$. The two solutions are shown in Fig. 1a, where the positive root range is $0 \leq v < \frac{1}{2}$ (solid line) and the negative root range is $-1 < v \leq 0$ (dashed line). The function is continuous where the two roots converge at $E/M = 1$, when $v = 0$. Likewise the shear modulus can be found as

$$G = \frac{M}{8}\left[\frac{E}{M} + 3 \mp \left(\frac{E^2}{M^2} - 10\frac{E}{M} + 9\right)^{1/2}\right]. \tag{7}$$

Substituting eqs. 6 and 7 into $2G(1-v) = M(1-2v)$ reveals that the positive root in $v$ is linked to the negative root in $G$, which is indicated by the $\mp$ sign in eq. 7. The ratio $G/M$ is plotted in Fig 1a with a solid line for the negative root and a dashed line for the positive root. There is a similar quadratic formula for the bulk modulus with a link to the signs of the roots in eqs. 6 and 7. However, in real materials the elastic constants are unique at any given state; e.g., there is only one bulk modulus at any given temperature and pressure. Since there must be a single value of $v$, $G$, and $B$ for any value of $E$ and $M$, only one set of roots is valid.

In biaxial loading $\sigma = \sigma_{11} = \sigma_{22}$, with all other stresses equal to zero, and $\varepsilon = \varepsilon_{11} = \varepsilon_{22}$. The biaxial elastic constant is[11]

$$H = \frac{\sigma}{\varepsilon}. \tag{8}$$

The constitutive stress-strain relations show that $H = E/(1-v)$, and since $E > 0$ and $-1 < v < \frac{1}{2}$, it follows that $H > 0$. Of course an expression between $H$ and any other two elastic constants may be derived. The quadratic relationships are of special interest. The biaxial modulus as a function of $E$ and $M$ is

$$H = \frac{E}{4}\left[5 - \frac{E}{M} \pm \left(\frac{E^2}{M^2} - 10\frac{E}{M} + 9\right)^{1/2}\right]. \qquad (9)$$

Equation 9 is plotted in Fig. 1a as the ratio $H/M$; the maximum occurs in the positive root at $E/M = 9/10$, where $H/M = 9/8$. In the same way $\nu$ as a function of $H$ and $M$ is found to be

$$\nu = \frac{M}{2H + 4M}\left[\frac{2H}{M} - 1 \pm \left(9 - 8\frac{H}{M}\right)^{1/2}\right]. \qquad (10)$$

The two roots of this expression converge at $H/M = 9/8$, where $\nu = 1/5$. Figure 1b shows how the maxima in eqs. 6 and 10 divide Poisson's ratio into three ranges: $-1 < \nu \leq 0$, $0 \leq \nu \leq 1/5$, and $1/5 \leq \nu < 1/2$, of which only one can be valid.

### III. RESULTS AND DISCUSSION

The data in Table 1, listing Poisson's ratio for isotropic samples of pure elements,[12–19] engineering alloys,[13,16,20–25] polymers,[26–31] and ceramics,[32–45] show that $1/5 \leq \nu < 1/2$ is consistent with experiment. The list is not exhaustive, but does provide a representative survey of 40 materials, encompassing the four major classes of solids. Note that data for certain materials were unavailable, so the table lists volumetric averages from single crystal measurements, assuming that grain boundaries do not affect the elasticity of the aggregate. Further data can be found in Simmons and Wang.[40] Table 1 includes newer materials, such as bulk metallic glass (vitreloy) and nanolaminate ceramic ($Ti_3SiC_2$). Within experimental error all substances lie within $1/5 \leq \nu < 1/2$.

Three of the materials in Table 1 (vitreloy, silicate glasses, and concrete) have variable composition; hence, Poisson's ratio varies. Figure 3 further explores $\nu$ for compositionally variable solids, plotting Poisson's ratio for 121 glasses grouped by chemical system[6]. Within the experimental scatter $\nu \geq 1/5$, with the exception of pure $SiO_2$ glass (fused quartz).

Well-characterized substances for which $\nu < 1/5$ are α-beryllium, diamond, boron nitride, fused quartz, α-cristobalite, and TiNb$_{24}$Zr$_4$Sn$_{7.9}$ (β-type titanium) alloy. These outliers may be separated into two categories, hard materials (Be, diamond, BN) and metastable materials with a large void fraction (SiO$_2$ glasses, cristobalite, titanium alloy). Auxetic substances such as α-cristobalite are included in this list, and are not distinct from other homogenous materials which do not obey $1/5 \leq \nu < 1/2$. Also there are certain foams which have negative $\nu$.[46] While classical elasticity has been applied to the aggregate behavior of foams[47], they are not included in this discussion because their properties are not fundamental but arise from cell geometry.[48]

For the hard materials, measurements of Poisson's ratio for α-beryllium range from 0.021 to 0.116[8]. Poisson's ratio for diamond is known more accurately, and for random aggregates is calculated to be 0.069[7]. Measurements of $\nu$ of vapor-deposited diamond are complicated by texture[49], and of sintered diamond by binder[50]; nevertheless, it appears that $\nu$ is less than $1/5$. Resonant ultrasound measurements of sintered cubic boron nitride have found $\nu \sim 0.14 - 0.18$,[50] which is somewhat larger than that predicted from volumetric averaging of the single crystal. Sintering BN to full density complicates the determination due to the sintering aid.

As shown in Fig. 2, $\nu$ for pure SiO$_2$ glass is in the range 0.15-0.16. Interestingly, it is possible to densify glasses, with the change in volume correlated to the inverse of Poisson's ratio. The volume change for fused quartz was large, 21%, which increased $\nu$ to 0.33.[51] Poisson's ratio for the low temperature form of cristobalite was found to be negative, which has been attributed to the rotation of the SiO$_4$ tetrahedra akin to the rotation of ribs in auxetic foams.[52] A titanium alloy with $\nu = 0.14$ appears to be due to a strain-induced matensitic transformation.[53] The resemblance of these materials with large atomic voids to that of foams with microstructural pores is striking. Likewise there is a similarity to lightweight concrete, wherein the mineral aggregate (such as haydite) contains a significant fraction of voids; Poisson's ratio of these materials have been measured to be less than $\frac{1}{5}$.[45]

## IV. CONCLUSIONS

The equations and their roots derived herein are general. While the analysis does not determine which of the three ranges of $\nu$ is valid, from experimental data it is clear that $\frac{1}{5} \leq \nu < \frac{1}{2}$ is the correct result. This range can be used to identify additional constraints on the elastic constants, such as $E/M \leq 9/10$. The two-parameter theory cannot describe materials with large void fractions (e.g., α-cristobalite), irreversible structural changes (e.g., titanium alloy), or extremely hard substances (e.g., diamond). The use of eq. 2 to interpret the behavior of such materials[6-8,49-52] is incorrect. This failure has not been apparent heretofore because a test of classical elasticity requires three independently measured elastic constants, which generally are not available. However, deviation from the range of $\nu$ identified herein can be taken as an indication of the incorrectness of an analysis employing the classical equations.

## ACKNOWLEDGEMENTS

We thank Drs. R.B. Bogoslovov and D.M. Fragiadakis for useful suggestions. This work was supported by the Office of Naval Research.

---

TABLE 1: Poisson's ratio of isotropic materials at room temperature.

| MATERIAL | POISSON'S RATIO | REF. | MATERIAL | POISSON'S RATIO | REF. |
|---|---|---|---|---|---|
| *Elements**  |  |  | *Polymers* ‡ |  |  |
| C (graphite) | 0.31 | 12 | Polystyrene | 0.34 | 26 |
| Mg | 0.291 | 13 | Polycarbonate | 0.42 | 27 |
| Si † | 0.22-0.23 | 14 | Polyvinyl chloride | 0.38 | 27 |
| Cr | 0.21 | 13,15 | Polymethyl methacrylate | 0.365-0.375 | 28 |
| Cu | 0.355 | 16 | Polyethylene terephthalate | 0.29 | 29 |
| Zn | 0.25 | 13 | Polytetrafluoroethylene | 0.41-0.42 | 30 |
| Ag | 0.36 | 17 | Natural rubber | 0.4999 | 31 |
| Sn (metal) | 0.357 | 13,18 | *Ceramics** |  |  |
| W | 0.28 | 13,16 | MgO | $0.18 \pm 0.03$ | 32 |
| Au | 0.45 | 16 | NaCl | 0.253 | 33 |
| Pb | 0.46 | 16 | CsCl | 0.266 | 34 |
| U | 0.23 | 19 | $CaF_2$ † | 0.283 | 35 |
| *Engineering Alloys* |  |  | $Al_2O_3$ | 0.231 | 36 |
| Low alloy carbon steel | 0.29-0.30 | 13 | TiN | 0.25 | 37 |
| 18-8 Stainless steel | 0.305 | 16 | $BaTiO_3$ | 0.27 | 38 |
| Grey cast iron | 0.26 | 20,21 | $LiNbO_3$ † | 0.25-0.26 | 39,40 |
| 70-30 Brass | 0.331 | 16 | $Ti_3SiC_2$ | 0.20 | 41 |
| Aluminum 6061-T6 | 0.33 | 22 | $B_2O_3$ glass | 0.30 | 42 |
| Bronze | 0.34 | 23 | $GeO_2$ glass | 0.20 | 43 |
| Titanium (dental alloy) | 0.30-0.31 | 24 | silicate glasses | 0.20-0.276 | 44 |
| Cu-Zr-Be glass (vitreloy) | 0.35-0.39 | 25 | concrete | 0.20-0.37 | 45 |

\* Measurements of aggregate polycrystalline samples, except where noted.

† Volume average of single crystal elastic constants.

‡ Neat materials.

FIG. 1. Quadratic elasticity expressions. (a) The two solutions of eqs. 6, 7, and 10, labeled $v$, $G/M$ and $H/M$. Solid lines show the positive roots of eqs. 6 and 10 and the negative root of eq. 7, and vice-versa for the dashed lines. (b) Roots of eqs. 6 and 10, labeled $E/M$ and $H/M$, with the three possible ranges of Poisson's ratio drawn with different line types.

FIG. 2. Poisson's ratio of 121 inorganic glasses. Data compiled in ref. 6. Uncertainty standard deviation ranges from ±0.003 to ±0.01; representative error bar shows ±0.01.

Figure 1.

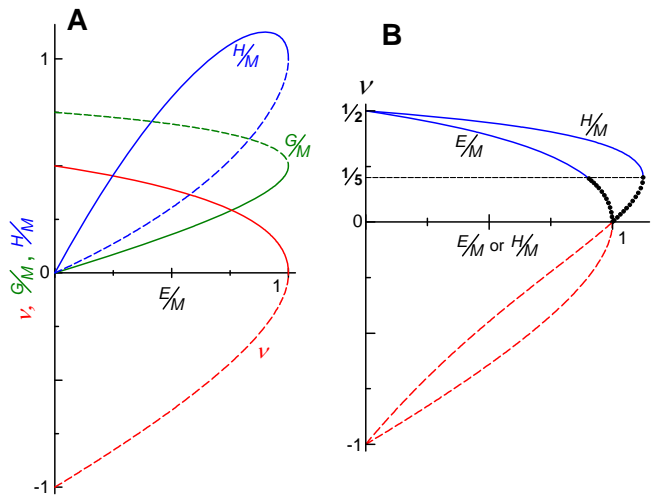

Figure 2.

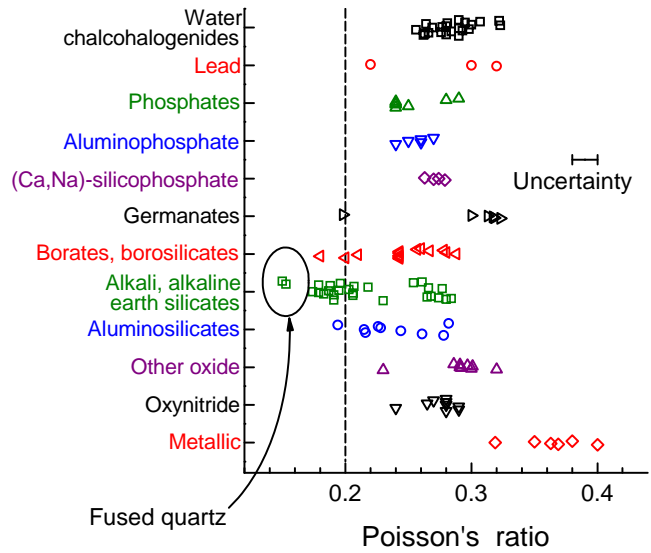